# $Q^2$ Dependence of the Azimuthal Asymmetry in Unpolarized Drell-Yan

**Elvio Di Salvo**

Dipartimento di Fisica and I.N.F.N. - Sez. Genova, Via Dodecaneso, 33

- 16146 Genova, Italy

**Abstract**

We study the azimuthal asymmetry of the unpolarized Drell-Yan in the framework of the T-odd functions. We find, on the basis of quite general arguments, that for $|\mathbf{q}_\perp| \ll Q$ such an asymmetry decreases as $Q^{-2}$, where $\mathbf{q}_\perp$ and $Q$ are respectively the transverse momentum and the center-of-mass energy of the muon pair. The experimental results support this conclusion.

# 1   Introduction

As is well-known, unpolarized Drell-Yan (DY) presents an azimuthal asymmetry. This was seen, for example, in reactions of the type[1-3]

$$\pi^- N \to \mu^+ \mu^- X, \qquad (1)$$

where $N$ is an unpolarized tungsten or deuterium target, which scatters a negative pion beam. Such an asymmetry was originally explained as a first order QCD correction effect[1]. Recently, however, it has been attributed[4, 5] to the quark polarization in unpolarized (or spinless) hadrons[6, 7]. Such a polarization reads

$$\mathbf{\Pi} = \frac{\mathbf{p}_\perp \times \mathbf{P}}{\mu_0 |\mathbf{P}|} h_1^\perp(x, \mathbf{p}_\perp^2). \qquad (2)$$

Here $x$ and $\mathbf{p}_\perp$ are, respectively, the Bjorken variable and the transverse momentum of the quark, $\mathbf{P}$ is the hadron momentum and $h_1^\perp$ a Lorentz invariant function introduced by Mulders and Tangerman[8] (MT). Lastly $\mu_0$ is a factor with the dimensions of a momentum, introduced for dimensional reasons and to be discussed below. Notice that this polarization, caused by a T-odd interference between two different amplitudes contributing to the process (1), is compatible with parity and time reversal invariance[6, 7, 9]. We limit our study to DY events such that

$$|\mathbf{q}_\perp| << Q, \qquad (3)$$

where $\mathbf{q}_\perp$ and $Q$ are, respectively, the transverse momentum and the center-of-mass energy of the muon pair. A problem we like to focus is the $Q^2$ dependence of the above azimuthal asymmetry. This behavior depends crucially on the factor $\mu_0$. MT assume it to be equal to the hadron rest mass, therefore they predict a substantial energy independence of the asymmetry. On the contrary, in this note we show that there are compelling reasons for defining differently that parameter; this implies that the azimuthal asymmetry decreases as $Q^{-2}$. As we shall see, data support such a conclusion.

The paper is organized as follows. In sect. 2 we give the general formulae for the DY cross section, illustrating the parameters which show deviations from the naive parton model. Sect. 3 is dedicated to the quark correlation matrix, containing "soft" information for the inclusive cross sections at high momentum transfers. In sect. 4 we determine the parameter $\mu_0$, either by comparison between the correlation matrix and the density matrix, or starting from a simple model. In sect. 5 we calculate the azimuthal asymmetry parameter $\nu$ (see formula (4) below) in terms of the polarizations $\Pi$ of the active quarks. Using the result of sect. 4, we obtain the $Q^2$ dependence of this asymmetry, which we compare with experimental results. Lastly we draw a short conclusion.

## 2  The unpolarized Drell-Yan cross section

The DY angular differential cross section is conventionally expressed as

$$\frac{1}{\sigma}\frac{d\sigma}{d\Omega} = \frac{3}{4\pi}\frac{1}{\lambda+3}(1+\lambda cos^2\theta + \mu sin2\theta cos\phi + \frac{1}{2}\nu sin^2\theta cos2\phi). \qquad (4)$$

Here $\Omega = (\theta, \phi)$, where $\theta$ and $\phi$ are respectively the polar and azimuthal angle of the momentum of the positive muon in a given center-of-mass frame of the muon pair. In particular, in the present paper we adopt the Collins-Soper (CS) frame[10], whose $z$-axis is along the bisector of the beam momentum and of the target momentum, while the $x$-axis is parallel to $\mathbf{q}_\perp$. Moreover $\lambda$, $\mu$ and $\nu$ are parameters, which are functions of the overall center-of-mass energy squared $s$, of $\mathbf{q}_\perp^2$, of $Q$ and of the longitudinal fractional momentum $x_F$ of the muon pair. On the left-hand-side, $d\sigma/d\Omega$ is a shorthand notation for

$$\frac{d\sigma}{d\Omega} \to \frac{d\sigma}{d\Omega dQ^2 dx_F d^2q_\perp}, \qquad (5)$$

where $x_F$ is the Feynman longitudinal fractional momentum of the muon pair with respect to the initial beam. In the naive DY model, where the parton transverse momentum and QCD corrections are neglected, one has $\lambda = 1$, $\mu = \nu = 0$. Therefore deviations of such parameters from the above naive predictions can be attributed to a nonvanishing transverse momentum of the partons inside the hadron or to gluon effects. In particular $\lambda \neq 1$ is due uniquely to the former cause. Since such deviations have been measured for all three parameters, it appears most natural to interpret them, as far as possible, as an effect of the quark transverse momentum, essential in the polarization (2). In particular, such an interpretation of the azimuthal asymmetry accounts rather well[4] for the observed large size of $\nu$ and small size of $\mu$[1, 2].

## 3 The quark correlation matrix

The correlation matrix is a very important tool for calculating inclusive reaction cross sections at high momentum transfers. Indeeed, in the framework of factorization[11], this matrix contains all information concerning the "soft" functions entering the theoretical expressions of such cross sections. For example, the DY cross section, which we are interested in, has the following convolutive expression:

$$\frac{d\sigma}{d\Omega dQ^2 dx_F d^2 q_\perp} = \frac{1}{(x_a + x_b)s}\frac{\alpha^2}{Q^4} L^{\mu\nu} W_{\mu\nu}. \tag{6}$$

Here $\alpha$ is the fine structure constant, $s$ the squared energy in the overall center-of-mass system and $x_a$ and $x_b$ the longitudinal fractional momenta of the active partons, with $x_F = x_a - x_b$. Moreover $L^{\mu\nu}$ and $W^{\mu\nu}$ are respectively the leptonic and hadronic tensor, i. e.,

$$L^{\mu\nu} = k^\mu \overline{k}^\nu + k^\nu \overline{k}^\mu - g^{\mu\nu} k \cdot \overline{k}, \tag{7}$$

$k$ and $\overline{k}$ being the four-momenta of the two leptons, and

$$W^{\mu\nu} = \frac{1}{6} \int d^2 p_\perp Tr\left[\Phi_A(x_a, \mathbf{p}_\perp)\gamma^\mu \Phi_B(x_b, \mathbf{q}_\perp - \mathbf{p}_\perp)\gamma^\nu\right]. \tag{8}$$

$\Phi_A$ and $\Phi_B$ are the correlation matrices of the active quark and antiquark, which belong to the two initial hadrons, denoted here respectively as $A$ and $B$. * The correlation matrix may be parametrized according to the Dirac algebra, taking into account Lorentz and parity invariance. On the contrary, we do not consider restrictions due to time reversal, since we admit the so-called T-odd functions, i. e., some particular interference terms, which, without violating any symmetry, change their sign under that transformation. Obviously, not all the Dirac components will contribute to a single process.

---

*For the sake of brevity, we omit the contribution of the annihilation between an antiquark of the hadron $A$ and a quark of the hadron $B$. However this does not affect our conclusion.

In order to determine $\mu_0$, it is necessary to consider a polarized nucleon, *e. g.*, a transversely polarized one. In this case the correlation matrix is parametrized as

$$\Phi = \Phi_\perp \simeq \Phi_{0a} + \Phi_{0b} + \Phi_1 + \Phi_O, \tag{9}$$

where

$$\begin{aligned}
\Phi_{0a} &= \frac{1}{2} p^+ \left( f_1 \slashed{n}_+ + \lambda_\perp g_{1T} \gamma_5 \slashed{n}_+ + \frac{1}{2} h_{1T} \gamma_5 [\slashed{S}, \slashed{n}_+] \right) \\
&+ \frac{1}{4\sqrt{2}} \lambda_\perp h_{1T}^\perp \gamma_5 [\slashed{p}_\perp, \slashed{n}_+], \tag{10}
\end{aligned}$$

$$\begin{aligned}
\Phi_{0b} &= \frac{1}{2} \left( f_1^\perp + \lambda_\perp g_T^\perp \gamma_5 \right) \slashed{p}_\perp \\
&+ \frac{1}{4} \lambda_\perp \left( h_T^\perp \gamma_5 [\slashed{S}, \slashed{p}_\perp] + h_T \mu_0 \gamma_5 [\slashed{n}_-, \slashed{n}_+] \right), \tag{11}
\end{aligned}$$

$$\Phi_1 = \frac{1}{2} M \left( e + g_T \gamma_5 \slashed{S} \right). \tag{12}$$

$$\Phi_O = \frac{1}{2\mu_0} \left[ f_{1T}^\perp \epsilon_{\mu\nu\rho\sigma} \gamma^\mu n_+^\nu p_\perp^\rho S^\sigma + i h_1^\perp \frac{1}{2} [\slashed{p}_\perp, \slashed{n}_+] \right] + \frac{M}{2P^+} i e_s \gamma_5 p_\perp \cdot S. \tag{13}$$

Here we have used the MT notations for the "soft" functions. Moreover we have set

$$\lambda_\perp = -S \cdot p_\perp / \mu_0, \qquad p_\perp = p - P \cdot p \frac{P}{M^2}, \tag{14}$$

$M$ being the nucleon rest mass, $P$ and $p$ the four-momenta of the nucleon and of the quark respectively, and $S$ the Pauli-Lubanski (PL) four-vector of the nucleon. Taking the $z$-axis along the nucleon momentum, it results $P \equiv (M/2P^+, P^+, \mathbf{0}_\perp)$, $p \equiv (p^-, p^+, \mathbf{p}_\perp))$ and $p_\perp \equiv (0, 0, \mathbf{p}_\perp)$. $n_\pm$ are lightlike vectors, such that $n_+ \cdot n_- = 1$ and whose space components are directed along the nucleon momentum.

The term $\Phi_O$ is T-odd. In particular, for an unpolarized nucleon, this reduces to

$$\Phi_O^U = \frac{i}{2} h_1^\perp [\slashed{p}_\perp, \slashed{n}_+], \tag{15}$$

which corresponds to the quark polarization $\mathbf{\Pi}$ (eq. (2)), as it is straightforward to check[6, 7]. It is important to stress that this polarization is due to a coherence effect, for example to one gluon exchange between the hadron $A$ and the active parton of the hadron $B$, or *vice-versa*[12]. This effect can be factorized only if the condition (3) is fulfilled[4]. Moreover the factor $\mu_0$ *must* be the same for all the functions involved, in order to normalize them appropriately. This factor was assumed to be equal to the rest mass of the hadron[8, 13]. As we shall see in a moment, this is not the most suitable choice, in order to interpret the functions contained in the correlation matrix as probability densities or interference terms.

## 4  Determining $\mu_0$

We follow two different procedures for deriving the correct expression of the parameter $\mu_0$. The first procedure consists in comparing the correlation matrix with the density matrix, to which $\Phi$ reduces for non-interacting quarks. The second procedure is based on a simple, but rather general, model, without any *ad hoc* assumptions.

### 4.1  The density matrix

The density matrix of a free, on-shell quark in a transversely polarized nucleon reads

$$\rho_\perp = \sum_{T=\pm 1/2} q_T(x, \mathbf{p}_\perp) \frac{1}{2}(\slashed{p} + m)(1 + 2T\gamma_5 \slashed{S}_q). \tag{16}$$

Here $m$ is the rest mass of the quark, such that $p^2 = m^2$. $2TS_q$ is the *quark* PL vector, with $S_q^2 = -1$. $q_T(x, \mathbf{p}_\perp)$ is the probability density of finding a quark with its spin aligned with ($T = 1/2$) or opposite to ($T = $ -1/2) the proton spin. For the sake of simpolicity, although not essential in our formulae, we consider a reference frame - called $\mathcal{P}$-frame from now on - where the nucleon momentum has a modulus $|\mathbf{P}| = \mathcal{P} >> M$. In appendix we show that in such a frame one has

$$\begin{aligned}\rho_\perp &= \frac{1}{2} q(x, \mathbf{p}_\perp^2)(\slashed{p} + m) \\ &+ \frac{1}{2} \delta q_\perp(x, \mathbf{p}_\perp) \gamma_5 \left\{ \frac{1}{2}[\slashed{S}, \slashed{p}] + \slashed{p}\sin\theta' - C_1 + mC_2 \right\} + O(\mathcal{P}^{-1}).\end{aligned} \tag{17}$$

Here we have set

$$q(x, \mathbf{p}_\perp^2) = \sum_{T=\pm 1/2} q_T(x, \mathbf{p}_\perp), \qquad \delta q_\perp(x, \mathbf{p}_\perp) = \sum_{T=\pm 1/2} 2T q_T(x, \mathbf{p}_\perp), \tag{18}$$

$$C_1 = E_q \frac{1}{2}[\slashed{n}'_+, \slashed{n}'_-]\sin\theta', \tag{19}$$

$$C_2 = \slashed{S} + \frac{1}{\sqrt{2}}\left\{\slashed{n}'_-\left(1 - \frac{m}{|\mathbf{p}|}\right) - \slashed{n}'_+ + \frac{1}{\sqrt{2}}[\slashed{n}'_+, \slashed{n}'_-]\right\}sin\theta' \qquad (20)$$

and

$$sin\theta' = -p_\perp \cdot S/|\mathbf{p}|. \qquad (21)$$

Lastly $n'_\pm$ are defined analogously to $n_\pm$, but with the space component along the quark momentum.

## 4.2 Comparison with the correlation matrix

We equate the coefficients of the independent Dirac operators in eqs. (17) and (9), taking into account the relation

$$p = \sqrt{2}x\mathcal{P}n_+ + p_\perp + O\left(\mathcal{P}^{-1}\right). \qquad (22)$$

We get, for a free, on-shell quark[14],

$$f_1 = f_1^\perp = q, \qquad \lambda_\perp h_T^\perp = sin\theta'\delta q_\perp, \qquad (23)$$

$$\lambda_\perp h_{1T}^\perp = (1 - \epsilon_1)sin\theta'\delta q_\perp, \qquad \mu_0\lambda_\perp h_T = (1 - \epsilon_1)sin\theta' E_q \delta q_\perp, \qquad (24)$$

$$\lambda_\perp g_{1T} = (1 - \epsilon_2)sin\theta'\delta q_\perp, \qquad \lambda_\perp g_T^\perp = (1 - \epsilon_3)sin\theta'\delta q_\perp. \qquad (25)$$

Here $\epsilon_1 = m/E_q$, $\epsilon_2 = m/x\mathcal{P}$ and $\epsilon_3 = m/2|\mathbf{p}|$ are the correction terms to the chiral limit, which are generally small for light quarks. The terms of order $O\left[(m^2 + \mathbf{p}_\perp^2)/\mathcal{P}^2\right]$ have been neglected. As regards $\mu_0$, we require the various functions to be normalized, in the chiral limit, like $\delta q_\perp$, which is a difference between two probability densities. Therefore we assume

$$\lambda_\perp = sin\theta', \qquad (26)$$

which, according to eqs. (21) and (14), implies

$$\mu_0 = |\mathbf{p}|. \qquad (27)$$

Two remarks are in order about $\mu_0$.

- This parameter is frame dependent, as well as the correlation matrix and the density matrix; however, in a specific reaction, it can be expressed in terms of

Lorentz invariant quantities. In particular, for DY, in the center-of-mass system we have

$$\mu_0 \simeq \frac{Q}{2}. \tag{28}$$

- $\mu_0$ is independent of the dynamics, therefore the presence of interactions among partons does not affect our result.

### 4.3 A model for $h_1^\perp$

A nucleon may be viewed as a bound state of the active quark with a set $X$ of spectator partons. In order to take into account coherence effects, we project the bound state onto scattering states with a fixed third component of the total angular momentum with respect to the nucleon momentum, $J_z$, and with a spin component $s = \pm 1/2$ of the quark along a given (axial) unit vector $\mathbf{s}$. For the sake of simplicity, following Brodsky et al.[12, 15] (see also [16, 17, 18]), we assume $X$ to have spin zero, moreover we choose a state with $J_z = 1/2$. Then

$$|J_z = 1/2; s; X\rangle = \alpha|\rightarrow, L_z = 0; s; X\rangle + \beta|\leftarrow, L_z = 1; s; X\rangle. \tag{29}$$

Here $\rightarrow (\leftarrow)$ and $L_z$ denote the components along $\mathbf{P}$, respectively, of the quark spin and orbital angular momentum, while $\alpha$ and $\beta$ are Clebsch-Gordan coefficients. Then the probability of finding a quark with $J_z = 1/2$ and spin component $s$ along $\mathbf{s}$, in a longitudinally polarized proton with a positive helicity, is

$$|\langle P, \Lambda = 1/2|J_z = 1/2; s; X\rangle|^2 = \alpha^2|\langle P, \Lambda = 1/2|\rightarrow, L_z = 0; s; X\rangle|^2$$
$$+ \beta^2|\langle P, \Lambda = 1/2|\leftarrow, L_z = 1; s; X\rangle|^2 + I, \tag{30}$$

$$I = 2\alpha\beta Re\left[\langle P, \Lambda = 1/2|\rightarrow, L_z = 0; s; X\rangle\langle(\leftarrow, L_z = 1; s; X)|P, \Lambda = 1/2\rangle\right]. \tag{31}$$

We are especially interested in the interference term $I$, which may be interpreted as a quark polarization, independent of the nucleon polarization. Indeed, eq. (30) implies that $2I$ is the difference between the probability of finding a quark with spin component $s$ along $\mathbf{s}$ and the probability that the spin of a quark along the same quantization axis be $-s$. After partial wave expansion, $I$ reads

$$I = 2\sum_{l,l'=0}^{\infty} Re\left[ie^{-i\phi_0} A_l B_{l'}^*\right] P_l(cos\theta_0)P_{l'}^1(cos\theta_0). \tag{32}$$

Here $A_l$ and $B_l$ are related to partial wave amplitudes; moreover $\theta_0$ and $\phi_0$ are respectively the polar and the azimuthal angle of the quark momentum, assuming the polar axis along $\mathbf{P}$ and, as the azimuthal plane, the one through $\mathbf{P}$ and $\mathbf{s}$. In the $\mathcal{P}$-frame one has

$$P_l(cos\theta_0) \sim 1, \qquad P_l^1(cos\theta_0) \sim \frac{|\mathbf{p}_\perp|}{x\mathcal{P}}. \tag{33}$$

Then eq. (32) yields

$$I \sim \frac{|\mathbf{p}_\perp|}{x\mathcal{P}} \left(U cos\phi_0 + V sin\phi_0\right), \tag{34}$$

where $U$ and $V$ are real functions of $x$ and $\mathbf{p}_\perp^2$, made up with $A_l$ and $B_l$. Since $\mathbf{s}$ is an axial vector, parity conservation implies $U = 0$. Therefore

$$I \sim \frac{|\mathbf{p}_\perp|}{x\mathcal{P}} V sin\phi_0 = \pm \frac{V}{x\mathcal{P}} |\mathbf{p}_\perp \times \mathbf{s}|, \tag{35}$$

where the $\pm$ sign depends on the sign of $sin\phi_0$. Therefore the interference term $I$ is T-odd. Moreover, setting $\mathbf{s} = \mathbf{P}/|\mathbf{P}|$, and comparing eq. (35) with eq. (2), we identify $h_1^\perp$ with $V$ and we get

$$\mu_0 = x\mathcal{P}. \tag{36}$$

Eqs. (36) predicts that the quark tranverse polarization in an unpolarized (or spinless) hadron decreases as $\mathcal{P}^{-1}$. But in the center-of-mass frame, for $Q >> M$ one has $x\mathcal{P} \simeq Q/2$, therefore we recover the result (28).

## 5  Azimuthal asymmetry

Inserting formulae (7) to (13) into eq. (6), the differential cross section reads, under the condition (3),

$$\frac{d\sigma}{d\Omega dQ^2 dx_F d^2 q_\perp} = \frac{1}{(x_a + x_b)s} \frac{\alpha^2}{12Q^2} \sum_f e_f^2 \left[(1 + cos^2\theta)F_f + sin^2\theta cos2\phi \frac{H_f}{\mu_0^2}\right] \tag{37}$$

Here the sum runs over all the light flavors and antiflavors, $u, d, s, \overline{u}, \overline{d}, \overline{s}$. Moreover, omitting the flavor indices, we have

$$F = \int d^2 p_\perp f_{1A}(x_a, \mathbf{p}_\perp^2) f_{1B}[x_a, (\mathbf{q}_\perp - \mathbf{p}_\perp)^2], \tag{38}$$

where $f_{1A}$ and $f_{1B}$ are the unpolarized densities of the active partons. Lastly

$$H = \int d^2 p_\perp h_{1A}^\perp(x_a, \mathbf{p}_\perp^2) h_{1B}^\perp[x_a, (\mathbf{q}_\perp - \mathbf{p}_\perp)^2] \mathcal{S}, \tag{39}$$

where

$$\mathcal{S} = [2\mathbf{p}_\perp \cdot \mathbf{n}(\mathbf{q}_\perp - \mathbf{p}_\perp) \cdot \mathbf{n} - \mathbf{p}_\perp \cdot (\mathbf{q}_\perp - \mathbf{p}_\perp)] \tag{40}$$

and $\mathbf{n} = \mathbf{q}_\perp/|\mathbf{q}_\perp|$. But eq. (39) implies that for $\mathbf{q}_\perp \to 0$ $H \propto \mathbf{q}_\perp^2$:

$$H = H_0 \mathbf{q}_\perp^2, \tag{41}$$

where $H_0 = H_0(x_a, x_b, \mathbf{q}_\perp^2)$ assumes a finite value for $\mathbf{q}_\perp = 0$. Comparing eq. (37) with eq. (4) yields

$$\nu = \frac{2}{\mu_0^2} \frac{\sum_f e_f^2 H_f}{\sum_f e_f^2 F_f}. \tag{42}$$

Inserting eqs.(41) and (28) into eq. (42), we get

$$\nu = A_0 \frac{\mathbf{q}_\perp^2}{Q^2} = A_0 \rho^2, \tag{43}$$

$A_0$ being a proportionality costant and $\rho = |\mathbf{q}_\perp|/Q$. Our prediction is compared with the experimental resuls of $\nu$ at different energies, both as a function of $Q$ at fixed $|\mathbf{q}_\perp| \ll Q$ (figs. 1 and 2, where also the MT assumption is tested) and as a function of $\rho$ (figs. 3 and 4).

Comparison of our result with data confirms that the value of $\mu_0$ deduced in sect. 4 (see eqs. (27) and (28)) has to be preferred to the one assumed by MT. In particular, such a result allows a very simple interpretation of the behavior of $\nu$ versus $\rho$, which would be quite difficult to fit with the MT ansatz.

# Appendix

Here we derive formula (17) for the density matrix of a quark. To this end we write formula (16):

$$\rho_\perp = \sum_{T=\pm 1/2} q_T(x, \mathbf{p}_\perp)\frac{1}{2}(\not{p} + m)(1 + 2T\gamma_5 \not{S}_q), \qquad (A.~1)$$

using the same notations as in subsect. 4.1. Moreover we define a *quark* rest frame, whose axes are parallel to those of the $\mathcal{P}$-frame, defined in subsect. 4.1. Here the Pauli-Lubanski vector of the quark results to be $S_q = S_q^{(0)} = S \equiv (0,0,1,0)$. Decomposing $S_q^{(0)} = S$ into a transverse and a longitudinal component with respect to the quark momentum, we get

$$S_q^{(0)} = S = \Sigma_\perp cos\theta' + \nu sin\theta'. \qquad (A.~2)$$

Here

$$sin\theta' = sin\theta sin\phi, \qquad sin\theta = \frac{|\mathbf{p}_\perp|}{|\mathbf{p}|}, \qquad sin\phi = \frac{-p_\perp \cdot S}{|\mathbf{p}_\perp|}, \qquad (A.~3)$$

$$\mathbf{p} \simeq (\mathbf{p}_\perp, x\mathcal{P}), \qquad \nu \equiv (0, \mathbf{t}), \qquad \Sigma_\perp \equiv (0, \mathbf{n}), \qquad (A.~4)$$

$$\mathbf{t} \equiv (sin\theta cos\phi, sin\theta sin\phi, cos\theta), \qquad (A.~5)$$

$$\mathbf{n} \equiv (cos\theta cos\phi, cos\theta sin\phi, -sin\theta). \qquad (A.~6)$$

In order to calculate $S_q$ in the $\mathcal{P}$-frame, we perform a boost along the quark momentum. This boost leaves $\Sigma_\perp$ invariant and transforms $\nu$ into $\tilde{p}/m$, where

$$\tilde{p} \equiv (|\mathbf{p}|, E_q \mathbf{t}), \qquad E_q = \sqrt{m^2 + \mathbf{p}^2}. \qquad (A.~7)$$

As a result we get

$$S_q = S + \left[\frac{p}{m} - (\delta + \nu)\right] sin\theta', \qquad (A.~8)$$

having defined

$$\delta = \frac{m}{\sqrt{2}x\mathcal{P}} n'_- \left[1 + O(\mathcal{P}^{-2})\right], \qquad n'_\pm \equiv \frac{1}{\sqrt{2}}(1, \pm\mathbf{t}). \qquad (A.~9)$$

Now we substitute eq. (A. 8) into eq. (A. 1), taking into account the definitions (A. 4) and (A. 9) of $\nu$ and $\delta$, and exploiting the relations $-\not{p}\not{S} = 1/2[\not{S},\not{p}] - p \cdot S$, $p \cdot S = p_\perp \cdot S$ and $\nu = \frac{1}{\sqrt{2}}(n'_+ - n'_-)$. As a result we get eq. (17).

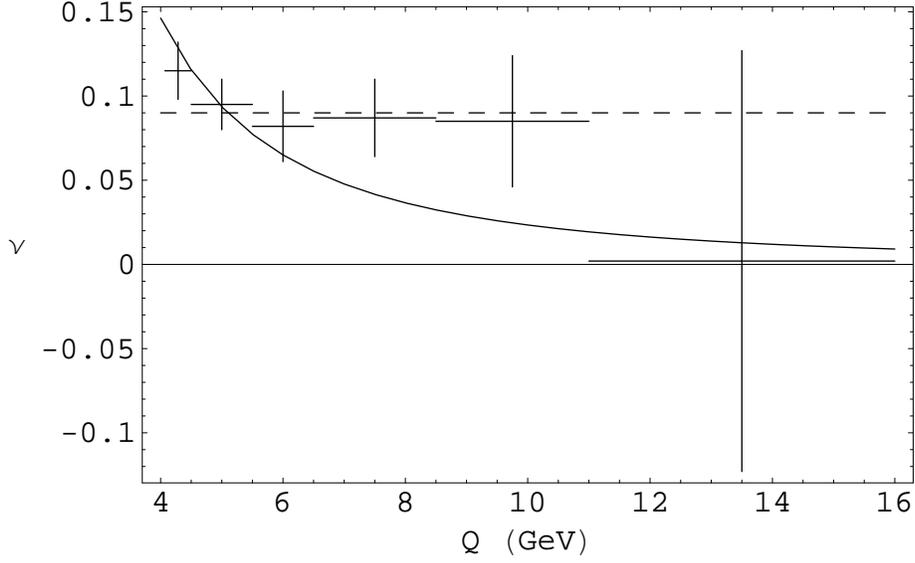

Figure 1: The behaviour of the asymmetry parameter $\nu$ vs $Q$, at constant $|\mathbf{q}_\perp| \ll Q$ and $\sqrt{s} = 19.1 \ GeV$. Data are taken from ref. 1 and fitted with formula $\nu = B_0/Q^2$, $B_0 = 2.34 \ GeV^2$. The dashed line corresponds to the MT ansatz.

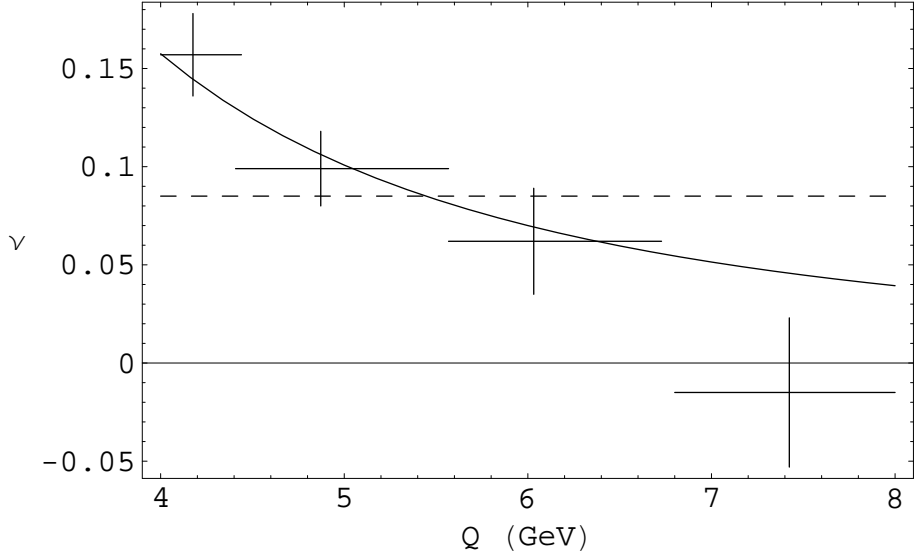

Figure 2: Same as fig. 1, $\sqrt{s} = 23.2\ GeV$, $B_0 = 2.52\ GeV^2$. Data are taken from ref. 2

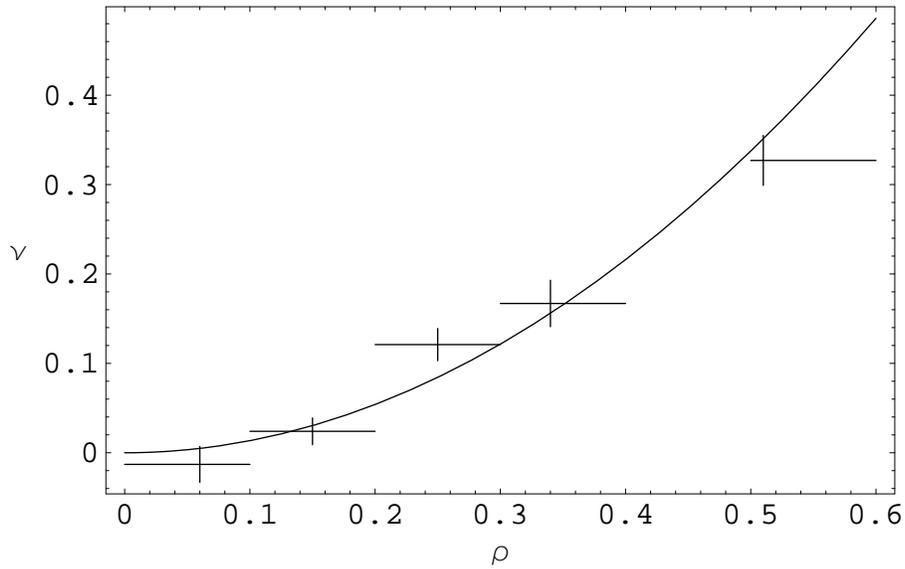

Figure 3: The behaviour of the asymmetry parameter $\nu$ vs $\rho = |\mathbf{q}_\perp|/Q$, $\sqrt{s} = 19.1\ GeV$. Data are taken from ref. 2 and fitted with formula $\nu = A_0\rho$, $A_0 = 1.35$

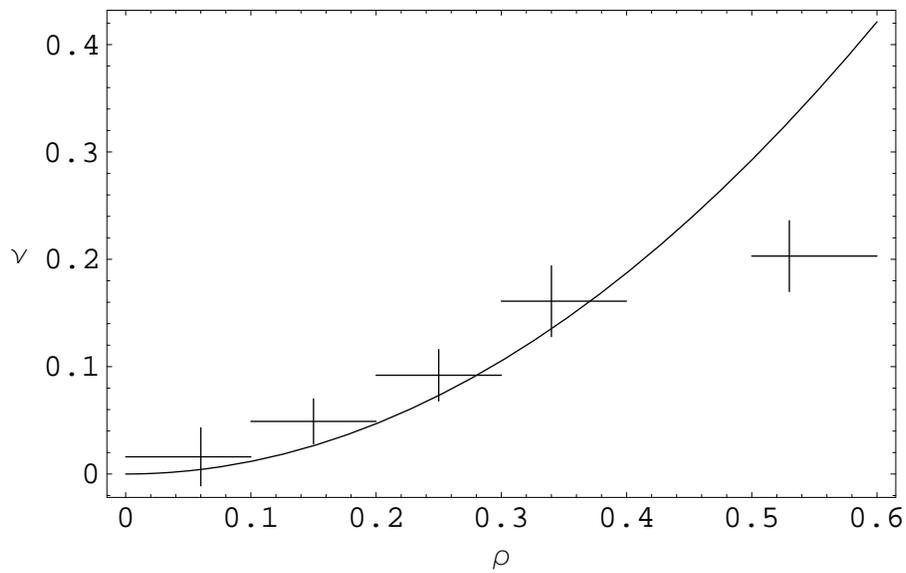

Figure 4: Same as fig. 3, $\sqrt{s} = 23.2\ GeV$. $A_0 = 1.17$